# Effect of the COVID-19 vaccine on preventive behaviors: Evidence from Japan

Eiji Yamamura[1*], Youki Kosaka[2], Yoshiro Tsutsui[3], Fumio Ohtake[4],

[1] Department of Economics, Seinan Gakuin University, Fukuoka, Japan

[2] Kyoto Economic College, Japan

[3] Kyoto-Bunkyo University, Japan

[4] Osaka University, Japan

# Abstract

Vaccination against the coronavirus disease 2019 (COVID-19) is a key measure to reduce the probability of getting infected with the disease. Accordingly, this might significantly change an individual's perception and decision-making in daily life. For instance, it is predicted that with widespread vaccination, individuals will exhibit less rigid preventive behaviors, such as staying at home, frequently washing hands, and wearing a mask. We observed the same individuals on a monthly basis for 18 months, from March 2020 (the early stage of the COVID-19 pandemic) to September 2021, in Japan to independently construct large sample panel data (N=54,007). Using the data, we compare the individuals' preventive behaviors before and after they got vaccinated; additionally, we compare their behaviors with those individuals who did not get vaccinated. Furthermore, we compare the effect of vaccination on the individuals less than or equal to 40 years of age with those greater than 40 years old. The major findings determined after controlling for individual characteristics using the fixed effects model and various factors are as follows. First, as opposed to the prediction, based on the whole sample, the vaccinated people were observed to stay at home and did not change their habits of frequently washing hands and wearing a mask. Second, using the sub-sample of individuals aged equal to or below 40, we find that the vaccinated people are more likely to go out. Third, the results obtained using a sample comprising people aged over 40 are similar to those obtained using the whole sample. Preventive behaviors are affecting oneself and generating externalities on others during this pandemic. Informal social norms motivate people to increase or maintain preventive behaviors even after being vaccinated in societies where such behaviors are not enforced.

# 1. Introduction

To reduce the transmission of the coronavirus disease 2019 (COVID-19), many countries implemented a lockdown. Individuals were obligated to follow the preventive behaviors enforced by the government; otherwise, they were penalized[1]. Economic activities were suspended, following restrictions on movements in daily life[2,3]. Subsequently, the lockdown significantly reduced the contact rate and spread of COVID-19[4–6].

However, lockdown restrictions result in a significant economic loss[7,8] and cause a detrimental impact on individuals' mental conditions[9–12]. Therefore, the Japanese government declared a "state of emergency" in which preventive behaviors are strongly required but not enforced. Even without the enforcement, Japanese individuals voluntarily exhibited preventive behaviors by staying at home, frequently washing their hands, and wearing a mask[13].

Various vaccines against the COVID-19 have been developed and distributed worldwide; vaccination was expected to act as a significant measure in the reduction of the spread of COVID-19 as well as depression among individuals caused by lockdowns. The number of newly reported cases of COVID-19 was observed to reduce in countries where the vaccines became rapidly pervasive[14]. Accordingly, the mental conditions of the vaccinated individuals improved[15].

Widespread vaccination is expected to promote economic activities because vaccinated people may not exhibit rigid preventive behaviors and adopt the lifestyle that they had before the spread of COVID-19[16–19]. However, Japanese individuals did not change their consumption behavior even after getting vaccinated; however, they are likely to increase their consumption after the eradication of COVID-19[20].

It is noteworthy to analyze the mechanism behind the unexpected consumption behavior in Japan. To this end, using monthly individual-level panel data, we investigate whether individuals' preventive behaviors change before and after getting vaccinated. Further, following the COVID-19 pandemic, individuals' reactions to changes in policies related to COVID-19 differ according to their situation [13]. There are time lags for the diffusion of vaccination among different generations in Japan. In a simulation study, determining the economic loss observed during the pandemic depends on whether the COVID-19 vaccine is allocated according to age groups[21]. Hence, we compare the difference in the effect of the uptake of COVID-19 vaccines between young and old generations, considering that they are exposed to different situations.

We find that people are more inclined to stay at home after being vaccinated based on the whole sample. Similar results are obtained when we use a sub-sample comprising individuals who are over 40 years old. In contrast, vaccinated individuals in a sub-sample of the younger generation, involving those aged less than or equal to 40, are less likely to stay at home and are more likely to go to work or school. These results vary from a similar study conducted in the U.K. where vaccination did not affect preventive behaviors[22,23]. This study offers new insights into the effect of vaccination on preventive behaviors in the context of the COVID-19 pandemic.

## 2. Materials and methods

### 2.1. Data collection

We commissioned the research company INTAGE to conduct an internet survey for this study, based on their experience and reliability. Participants registered with INTAGE were recruited for the study. The sampling method was designed to gather a representative sample of the Japanese population in terms of gender, age, educational background, and residential area. Japanese citizens aged from 16-79 were selected for the survey. Internet surveys were conducted repeatedly for 15 separate times ("waves") almost every month with the same individuals to construct the panel data. However, the survey could not be conducted for the exceptional period between July 2020 and September 2020 because of a shortage of research funds. We resumed the surveys after receiving additional research funds from October 2020.

The first wave of queries was conducted from March 13, 2020 to March 16, 2020, recording 4,359 observations with a response rate of 54.7 %. Respondents from the first wave were targeted in the subsequent waves to record how the same respondent changed their perceptions and behaviors during the COVID-19 pandemic. During the study period, some of the respondents quit taking the surveys, while some did not take all the surveys. The total number of observations used in this study is 54,007.

## 2.2. Ethical issues

Our study was performed following the relevant guidelines and regulations. The ethics committee of Osaka University approved all survey procedures, and informed consent was obtained from all participants.

All survey participants provided their consent to participate in the anonymous online survey. After being informed about the purpose of the study and their right to quit the survey, participants agreed to participate. The completion of the entire questionnaire was considered to indicate the participants' consent.

## 2.3. Measurements

Table 1 contains a description of the variables, their mean values, and standard deviations. The survey questionnaire contained basic questions about demographics such as age, gender, and educational background. Fifteen waves were conducted from March 2020 to September 2021. As the main variables, the respondents were asked questions concerning preventive behaviors as follows:

"Within a week, to what degree have you practiced the following behaviors? Please answer based on a scale of 1 (I have not practiced this behavior at all) to 5 (I have completely practiced this behavior)."

(1) Staying indoors,

(2) Not going out to the workplace (or school),

(3) Not going out to the events or travel,

(4) Washing my hands carefully,

(5) Wearing a mask.

The answers to these questions served as proxies for the following variables for preventive behaviors: staying indoors, not going out for work, not participating in leisure activities outside home, frequently washing hands, and wearing masks. Larger values indicate that respondents are more likely to engage in preventive behaviors. Staying indoors generally captures the degree of stay (not going out) at home. For more specific behaviors, we asked the type of voluntary restraint about going out. Not going out for work captures the degree of avoiding going out to work or school. Not participating in leisure activities outside home captures the degree of avoiding going out to events or travel. In the case of the former, preventive behavior depends on the condition of the workplace or school. Hence, there is a possibility that respondents are obliged to go to work or school. The latter is more likely to depend on an individual's decision-making. Further, we asked about the subjective probability of contacting COVID-19 and their perception of the severity of COVID-19.

Table 1 suggests that the mean values of staying indoors and not going out for work are 2.91 and 2.94, respectively. Meanwhile, the value of not participating in leisure activities outside home is 4.12. This means that people are more likely to go to work or school than to leisure. This suggests that events or travel are considered less essential than work or school. Staying indoors consists of both essential and non-essential components. Overall, not going out for work is a more critical factor of the degree of staying indoors. Not going out for work is determined not by an individual's will but by instruction from the workplace or school. Like refraining from leisure, the mean values of washing hands and wearing masks are slightly larger than 4. The reason is that washing hands and wearing masks are likely to depend on an individual's will.

We also asked the respondents whether they took the first shot of the vaccine against COVID-19 and asked whether they had completed the second vaccine shot. In Japan, vaccination began in February 2021[24]. During this period, the initial group to get the short was strictly limited to health workers before expanding its inoculation program. Vaccination for the general older people aged 65 and over has been implemented from April 2021, and then, 75 % of older people have been vaccinated in July 2021[25]. In addition, the government has started COVID-19 vaccination programs at workplaces and campuses where workers and students can get vaccinations from June[26].

The 10th wave survey was conducted directly after February 2021. In the sample used in this study, respondents who received the shot appeared from the 12th wave conducted in May 2021. As for the dummies for vaccination, the mean values of Vaccine second_1, Vaccine second_2, Vaccine second_3, and Vaccine second_4 are 0.03, 0.02, 0.01, and 0.001, respectively. This means that, in the whole sample, people who received the second shot at the survey time accounted for 3%. People who received the second shot last month, two months ago, and three months ago were 2 %, 1 %, and only 0.1%, respectively. The whole sample covered first-eleventh waves where nobody received the shot, and so percentages are very low. The vaccine was distributed to health care workers and older people, and others only from June. Therefore, the percentage declines as people who received the second shot earlier.

**Table 1. Definitions of key variables and their basic statistics.**

| Variables | Definition | Mean | s.d. |
|---|---|---|---|
| Staying indoors | In last week, how have you achieved "not going out of home?" Please choose from 5 choices.<br>1 (not completed at all) to 5 (completely achieved). | 2.91 | 1.25 |
| Not going out for work | In last week, how have you achieved "not going out to work (or school)?" Please choose from 5 choices.<br>1 (not completed at all) to 5 (completely achieved). | 2.94 | 1.73 |
| Not participating in leisure activities outside home | In last week, how have you achieved "not going out to events or travel?" Please choose from 5 choices.<br>1 (not completed at all) to 5 (completely achieved). | 4.12 | 1.18 |
| Washing hands | In last week, how have you achieved "washing your hands?" Please choose from 5 choices.<br>1 (not completed at all) to 5 (completely achieved). | 4.14 | 0.95 |
| Wearing mask | In last week, how have you achieved "wearing a mask?" Please choose from 5 choices.<br>1 (not completed at all) to 5 (completely achieved). | 4.41 | 1.05 |
| Vaccine First | Did you take the first shot (but not yet the second one)?<br>1 (Yes) or 0 (No) | 0.03 | 0.17 |
| Vaccine second | Did you take the second shot?<br>1 (Yes) or 0 (No) | 0.06 | 0.24 |
| Vaccine second_1 | Did you take the second shot in this month?<br>1 (Yes) or 0 (No) | 0.03 | 0.18 |
| Vaccine second_2 | Did you take the second shot last month?<br>1 (Yes) or 0 (No) | 0.02 | 0.14 |
| Vaccine second_3 | Did you take the second shot two months ago?<br>1 (Yes) or 0 (No) | 0.01 | 0.08 |
| Vaccine second_4 | Did you take the second shot three months ago?<br>1 (Yes) or 0 (No) | 0.001 | 0.04 |
| Probability COVID19 | What percentage do you think the probability of your taking the COVID-19?<br>0 to 100 (%) | 20.4 | 22.3 |
| Severity COVID19 | How serious are your symptoms if you are infected with the novel coronavirus? Choose from 6 choices.<br>1 (very small influence) to 6 (death) | 3.57 | 1.21 |
| Emergency | Areas where respondents reside are under the state of emergency.<br>1 (Yes) or 0 (No) | 0.29 | 0.45 |
| Age | Ages | 48.7 | 17.3 |
| Male | It takes 1 if respondent is male, otherwise 0. | 0.50 | 0.50 |
| University | It takes 1 if respondent graduated from university, otherwise 0. | 0.43 | 0.49 |

To closely check the change in the vaccination rate, Table 2 shows the percentages of vaccinated people in each wave. In contrast to the dummy for vaccination, Table 1 indicates the aggregated values containing both the first vaccinated people and the second vaccinated people regardless of vaccination time point. Therefore, the percentage of vaccinated individuals is expected to increase over time. Consistent with this inference, Table 2 indicates that the percentage of vaccinated people rapidly increased from 8.2% in May 2021 to 64.2% at the beginning of September in the sample. This rate is similar to that of 65.2 % in September in a country-wide sample[27]. As for the sub-sample of people over 40 years, the rate increased from 9.1% in May 2021 to 72.3%, almost two times higher than the sub-sample of younger people in each wave. Thus, the data of this study reflects the real situation of Japan.

**Table 2. Percentage of those who took the COVID-19 vaccine.**

| Waves | Dates | All % | Age>40 % | Age<=40 % |
|---|---|---|---|---|
| 1 | March 13–16, 2020 | 0 | 0 | 0 |
| 2 | March 27–30, 2020 | 0 | 0 | 0 |
| 3 | Apr. 10–13, 2020 | 0 | 0 | 0 |
| 4 | May 8–11, 2020 | 0 | 0 | 0 |
| 5 | June 12–15, 2020 | 0 | 0 | 0 |
| 6 | Oct 23–28, 2020 | 0 | 0 | 0 |
| 7 | Dec 4–8, 2020 | 0 | 0 | 0 |

| 8  | Jan. 15–19, 2021    | 0    | 0    | 0    |
| 9  | Feb. 17–22, 2021    | 0    | 0    | 0    |
| 10 | Mar. 24–29, 2021    | 0    | 0    | 0    |
| 11 | Apr. 23–26, 2021    | 0    | 0    | 0    |
| 12 | May 28–31, 2021     | 8.2  | 9.1  | 5.4  |
| 13 | June 25–30, 2021    | 25.1 | 30.7 | 7.8  |
| 14 | July 30–Aug 4, 2021 | 50.0 | 58.3 | 23.8.|
| 15 | Aug 27–Sep. 1, 2021 | 64.2 | 72.3 | 39.5 |

Note: We did not distinguish respondents who took only the first shot from those who took the second shot.

Fig 1 illustrates the change in five preventive behaviors from the first to the fifteenth waves by dividing the sample into vaccinated and non-vaccinated groups. Fig 1 covers the periods before and after vaccine distribution. Therefore, nobody was vaccinated from the first to the eleventh waves, where the left part of the vertical line is shown in Fig 1. In this study, people who were vaccinated during any period were included in the vaccinated group. Furthermore, we did not distinguish people who received the second shot from those who only received the first shot. For instance, one who was first vaccinated in the fifteenth wave was included in the vaccinated group. Thus, Fig 1 indicates how people who did not intend to be vaccinated behave differently from vaccinated people from the period when the vaccine was not distributed.

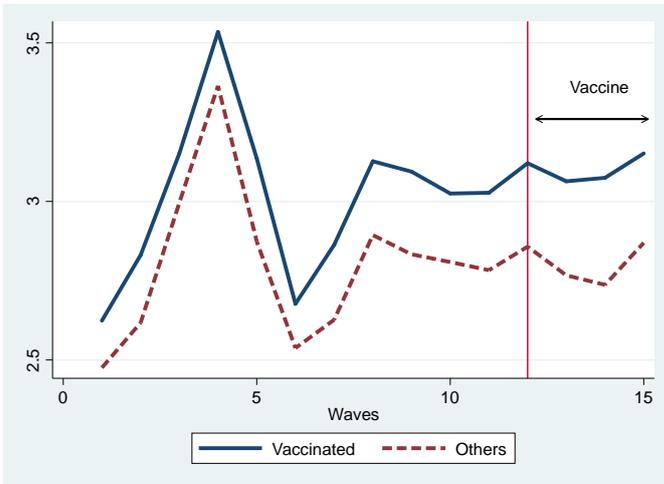

a) Staying indoors

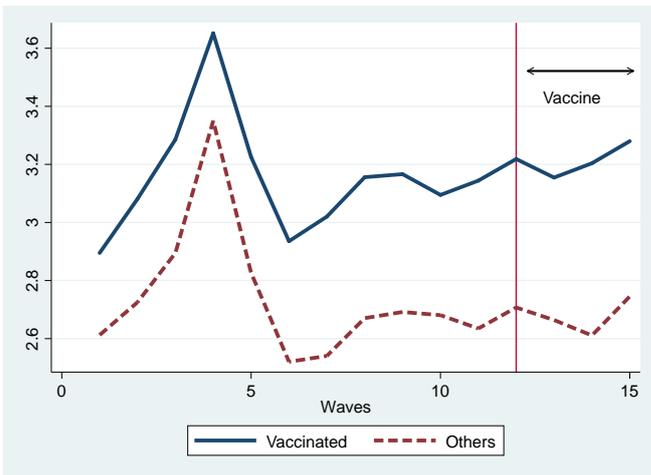

b) Not going out for work

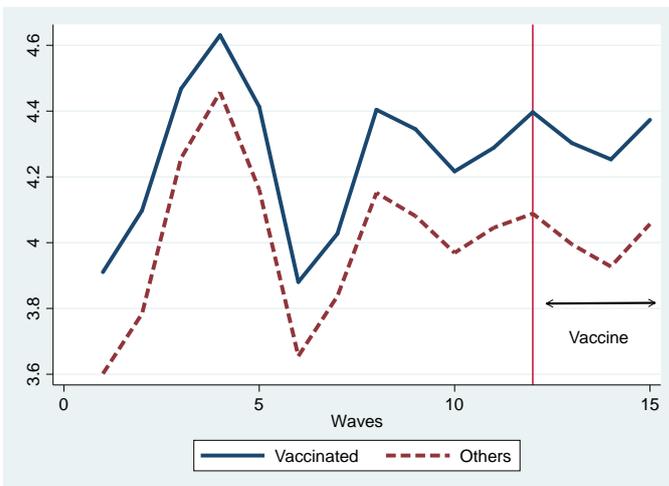

c) Not going to leisure

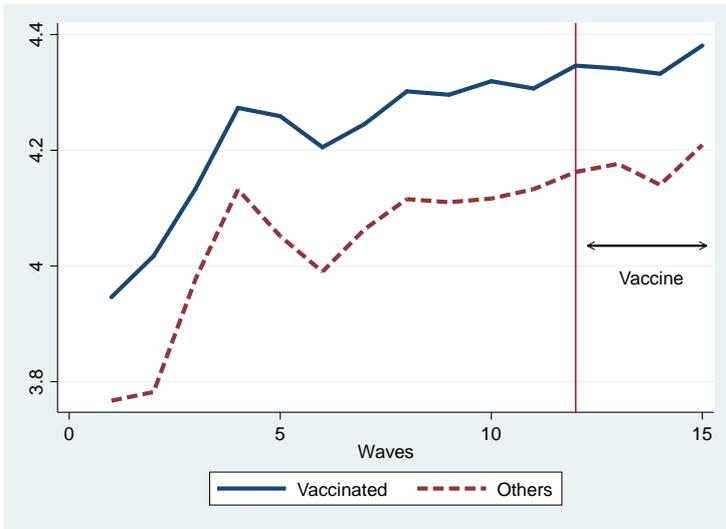

d) Washing hands

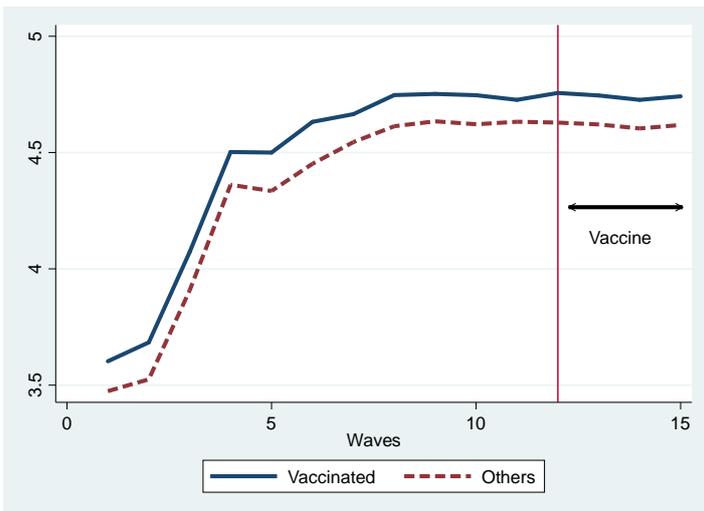

e) Wearing mask

Fig. 1.

Fig 1a indicates that the vaccinated group was more likely to stay at home than the non-vaccinated group throughout the studied period. The trend of both groups is similar. At the first declaration of a state of emergency in all parts of Japan from third to fourth waves, people drastically came to stay at home. Then, after calling the first declaration, the level of staying at home declined to the level before the declaration. Later, the state of emergency was declared and called off repeatedly four times. In response to it, the level of staying at home increased but not reached the peak level during the first declaration. Then, the level was more stable in 2021 than in 2020. However, we should notice that the gap of the behavior increased especially after the eighth wave and so after entering 2021. Similar tendencies were observed for Figs 1b and c for not going out for work and not participating in leisure activities outside home.

Turning to Figs 1d and e presenting changes in washing hands and wearing masks, similar to Figs 1a-c, the vaccinated group showed consistently higher levels than the non-vaccinated group. However, the gap of washing hands is larger than that of wearing mask. Wearing mask behavior is motivated by self-regarding risk preferences and other-regarding concerns[28–32]. That is, the effect of interpersonal interaction possibly reduces the gap in wearing masks between groups.

Compared to Figs 1a- c, there is a remarkable difference in the trends in Figs 1d and e. The level of washing hands and wearing masks almost constantly rose, indicating that people became more inclined to wash hands and wear a mask even after calling for a state of emergency. This is consistent with the fact that people came to wash their hands in response to the 2009 influenza pandemic, and their habits persisted over the years[33].

As a whole, in Fig 1, we do not observe the effect of vaccination by comparing before and after the distribution of the COVID-19 vaccine. Observations in Fig 1 are only the change in mean values, and thus various factors that influence preventive behaviors are not controlled. To closely examine the effects of vaccination, we examined the fixed effects regression model.

## 2.4. Methods

We used a fixed effects model regression to control the time-invariant individual fixed effects. The estimated function takes the following form:

$Y_{it} = \alpha_1 \text{ Vaccine First}_{it} + \alpha_2 \text{ Vaccine second\_1}_{it} + \alpha_3 \text{ Vaccine second\_2}_{it} + \alpha_4 \text{ Vaccine second\_3}_{it} + \alpha_5 \text{ Vaccine second\_4}_{it} + \alpha_6 \text{ Probability COVID19}_{it} + \alpha_7 \text{ Severity COVID19}_{it} + \alpha_8 \text{ Emergency}_{it} + k_t + m_i + u_{itg}$,

In this formula, $Y_{itp}$ represents the dependent variable for individual i and wave t. Y is preventive behaviors captured by five proxy variables: staying indoors, not going out for work, not participating in leisure activities outside home, washing hands, and wearing masks. Hence, in the same specification, we conduct five estimations separately. The regression parameters are denoted as α. The error term is denoted as *u*.

$k_t$ represents the effects of different time points, controlled by 14 wave dummies, where the first wave is the reference group. Various shocks occurred simultaneously throughout Japan at each time point. Wave dummies are included to control for it. The estimation method is the fixed effects (FE) model, and the time-invariant individual-level fixed effects are represented by $m_i$. This means that the model controls various individual characteristics that do not change even if time has passed. Hence, sex, educational background, and various factors are controlled. During the study period, respondents' ages increased only by one year, and the timing of change of ages depended on their birthday. Therefore, the variation in ages reflects that of birthdays by the FE model. Hence, age is not included in the model, even though the results do not change if age is included in the model.

Key independent variables are dummies for vaccination; Vaccine First controls the effect of the first shot. The vaccine was developed by various pharmaceutical companies. However, the Japanese government has approved only Pfizer and BioNTech vaccines. The first vaccinated persons are obliged to take the second shot within a month to make the vaccine effective. This rule applied to Pfizer and BioNTech vaccines. Then, the effect of the second shot should be estimated separately. Further, it is valuable to investigate how the effect of the vaccine on preventive behaviors changes over time. For this purpose, we included Vaccine second_1, Vaccine second_2, Vaccine second_3, and Vaccine second_4.

As control variables, in Japan, declarations of a state of emergency significantly affected behaviors[13,34]. However, the timing of the declarations differed according to the area. Therefore, the effect of the declaration cannot be captured by wave dummies. Hence, we include Emergency to control the effect of the emergency. Further, the subjective perception of COVID-19 is expected to influence preventive behavior. For instance, people are more likely to be cautious about COVID-19 if they consider the probability of taking COVID-19 to be higher and the damage of COVID-19 to be larger[35,36]. To control for this, we include Probability COVID19 and Severity COVID19. Although their results were not reported, we also controlled for the following factors: the number of persons infected with COVID-19 and deaths caused by COVID-19 in residential areas at each time point. Subjective emotions such as anxiety, fear, and anger are also controlled.

The motivation for getting a shot of vaccination depends on age[30,37]. The effect of vaccination is thought to differ according to the situation where people are confronted. In addition to estimation using the whole sample, we conduct estimations by dividing the sample into two sub-samples for young (below 40 years) and old generations (equivalent or over 40 years).

# 3. Results

## 3.1. Full sample estimations

Table 3 shows the estimation results of the FE model using the entire sample. We begin by examining the key variables of vaccination dummies. Except for column (5), where the wearing mask is the dependent variable, the coefficients of the vaccination dummies show a positive sign in most cases. Vaccine First is statistically significant only in columns (1) and (4), and its statistical significance is not at the 1% level. Further, Vaccine second_1 and Vaccine second_2 are statistically significant at the 1% level in most cases of columns (1)-(3), whereas Vaccine second_3 and Vaccine second_4 are not significant in any columns. Values of coefficients of Vaccine second_1 are approximately 0.07- 0.99 in columns (1) and (3). This can be interpreted as follows: in comparison with the non-vaccinated people, vaccinated people are more likely to stay home, not to go to work, or to leisure by 0.070-0.099 points on a 5-point scale. Turning to Vaccine second_2: their values increased to 0.106-0.123.

As a whole, these imply that people who completed the second shot come to stay at home and do not go out to work, school, or leisure. This tendency is observed in the month when they received the second shot and the next month. In particular, the effect is larger in the next month. However, this effect disappeared.

Before the estimation, we infer that the vaccination leads people to go out because COVID-19 is less likely to have a detrimental effect on vaccinated people than before vaccination. Our findings contradict this. After vaccination, some side effects are normal and expected, including pain, swelling, and redness on the arm where the vaccine was received, chills, mild fever, tiredness, headaches, joint pain, or muscle ache[38]. This possibly reduces the incentive to go out. However, the side effects disappear within a few days. Then, the side effect may affect one's ability to perform daily activities for a few days if one experiences the effect[38]. Thus, the side effect would not be the reason for the increase in staying at home in the following months.

In our interpretation, social norms to promote preventive behaviors were formed through the experience of COVID-19. After being vaccinated, there is an instruction from the experts "it is important that you continue to follow preventive measures after being vaccinated. This is because COVID-19 vaccines have proven effective at stopping people from developing the virus, but we do not yet know whether they prevent people from passing the infection onto others."[38] The instruction is considered as "nudge" to influence human behaviors[39–41]. Social media exposure to COVID-19 information influences the adoption of preventive attitudes and behaviors by shaping risk perception[42]. Arguably, the instruction after vaccination contributes to forming social norms through media.

People would have a sense of having done wrong if they do not follow the norms. Alternatively, vaccinated people predict to be punished if they break the norm. Especially at the early stage of vaccine distribution, the supply of vaccines is short, so only health care workers and adult people can receive the shot. Furthermore, it is challenging to make reservations for vaccination. Hence, highly advantageous vaccinated individuals are very small. They would be seriously criticized or take a bashing if they go out. If vaccinated people derive the inference, they refrain from going out.

Hence, the norms become more effective for vaccinated people because they are less likely to obey them. The gap in preventive behaviors between the vaccinated and the non-vaccinated people returned to the level before vaccination but did not decrease, although two or three months have passed. Concerning washing hands and wearing masks, dummies for vaccination did not show a significant negative sign. Therefore, vaccination did not hamper vaccinated people's preventive behaviors.

The model specification shows that subjective perception about COVID-19 is controlled by Probability COVID19 and Severity COVID19. In particular, the coefficient of Severity COVID19 exhibits a positive sign and is statistically significant at the 1% level in all estimations. This is consistent with the inference that people are more likely to exhibit preventive behaviors if they consider the damage of COVID-19 to be larger.

In most cases, wave dummies show a positive sign and statistical significance at the 1% level except for *Wave 6*. This suggests that people are more likely to display preventive behaviors than the first wave when COVID-19 arrived in Japan and did not spread so seriously. At the sixth wave, as illustrated in Fig 1, the level of preventive behaviors temporarily returned to the level in the early stage of the first wave when the first declaration of the state of emergency terminated. A significant positive sign of Emergency is observed for columns (2) and (3), which is reasonable because people are strongly required not to go out.

**Table 3. FE model. Dependent variables are preventive behaviors.**

|  | (1) Staying indoors | (2) Not going out for work | (3) Not participating in leisure activities outside home | (4) Washing hands | (5) Wearing mask |
|---|---|---|---|---|---|
| Vaccine First | 0.057** (0.02) | 0.032 (0.02) | 0.027 (0.02) | 0.026* (0.01) | −0.001 (0.02) |
| Vaccine second_1 | 0.099*** (0.02) | 0.070** (0.03) | 0.077*** (0.02) | 0.006 (0.02) | −0.006 (0.02) |
| Vaccine second_2 | 0.123*** (0.03) | 0.123*** (0.04) | 0.106*** (0.03) | −0.012 (0.02) | −0.0003 (0.02) |
| Vaccine second_3 | 0.097 (0.06) | 0.092 (0.06) | 0.018 (0.06) | 0.035 (0.03) | −0.027 (0.03) |
| Vaccine second_4 | 0.014 (0.17) | −0.019 (0.07) | −0.106 (0.11) | −0.018 (0.06) | −0.040 (0.06) |
| Probability COVID19 | −0.291 (0.38) | −0.001 (0.001) | 0.103 (0.21) | 0.428* (0.25) | −0.472 (0.36) |
| Severity COVID19 | 0.016*** (0.004) | 0.017* (0.01) | 0.036*** (0.01) | 0.018*** (0.005) | 0.033*** (0.01) |
| Emergency | 0.022 (0.02) | 0.034** (0.01) | 0.047*** (0.01) | −0.001 (0.01) | 0.013 (0.01) |
| Wave 1 | | | <Default> | | |
| Wave 2 | 0.126*** | 0.092*** | 0.170*** | 0.043** | 0.047*** |
| | (0.02) | (0.02) | (0.03) | (0.02) | (0.02) |
| Wave 3 | 0.446*** | 0.273*** | 0.516*** | 0.177*** | 0.386*** |
| | (0.04) | (0.04) | (0.03) | (0.02) | (0.03) |
| Wave 4 | 0.829*** | 0.687*** | 0.698*** | 0.329*** | 0.833*** |
| | (0.05) | (0.05) | (0.04) | (0.02) | (0.04) |
| Wave 5 | 0.435*** | 0.269*** | 0.517*** | 0.289*** | 0.862*** |
| | (0.04) | (0.04) | (0.03) | (0.01) | (0.03) |
| Wave 6 | 0.052 | −0.010 | 0.025 | 0.237*** | 1.010 |
| | (0.03) | (0.04) | (0.03) | (0.02) | (0.03) |
| Wave 7 | 0.161*** | 0.017 | 0.157*** | 0.267*** | 1.061*** |
| | (0.03) | (0.03) | (0.02) | (0.02) | (0.03) |
| Wave 8 | 0.389*** | 0.141*** | 0.458*** | 0.317*** | 1.122*** |
| | (0.04) | (0.04) | (0.03) | (0.02) | (0.04) |
| Wave 9 | 0.368*** | 0.153*** | 0.417*** | 0.319*** | 1.146*** |
| | (0.03) | (0.03) | (0.03) | (0.02) | (0.04) |
| Wave 10 | 0.344*** | 0.140*** | 0.323*** | 0.339*** | 1.133*** |
| | (0.04) | (0.03) | (0.03) | (0.02) | (0.03) |
| Wave 11 | 0.304*** | 0.132*** | 0.373*** | 0.336*** | 1.126*** |
| | (0.03) | (0.03) | (0.03) | (0.02) | (0.04) |
| Wave 12 | 0.375*** | 0.195*** | 0.442*** | 0.363*** | 1.139*** |
| | (0.03) | (0.03) | (0.03) | (0.02) | (0.04) |
| Wave 13 | 0.309*** | 0.141*** | 0.365*** | 0.372*** | 1.132*** |
| | (0.03) | (0.03) | (0.03) | (0.02) | (0.04) |
| Wave 14 | 0.282*** | 0.156*** | 0.278*** | 0.355*** | 1.111*** |
| | (0.04) | (0.04) | (0.04) | (0.02) | (0.04) |
| Wave 15 | 0.346*** | 0.231*** | 0.384*** | 0.413*** | 1.135*** |
| | (0.05) | (0.04) | (0.04) | (0.03) | (0.04) |
| Adj $R^2$ | 0.37 | 0.65 | 0.37 | 0.62 | 0.49 |
| Obs. | 54,007 | 54,007 | 54,007 | 54,007 | 54,007 |

**Note**: Numbers within parentheses are robust standard errors clustered on residential prefectures. For convenience, the coefficient of Probability COVID19 is multiplied by 1000. The model includes the number of deaths and infected persons in residential prefectures at the time of surveys and proxied for mental conditions such as fear, anxiety, and anger. However, its results are not reported. Are included, although the results are not reported.

***$p<0.01$

**$p<0.05$

*$p<0.10$

Table 4 presents different specifications where a second shot dummy is used to examine the effect of the second shot vaccination instead of using four dummies to capture the timing of the second shot. In Table 4, we only pay attention to whether respondents completed the second shot. We report the key variables, although the set of control variables are equivalent to Table 3. Results are similar to Table 3. The significant positive sign of Vaccine second was observed in columns (1)-(3), but not in columns (4) and (5). Its absolute values of coefficient and statistical significance are larger for Vaccine second than Vaccine First.

**Table 4. FE model. Dependent variables are preventive behaviors.**

|  | (1) Staying indoors | (2) Not going out for work | (3) Not participating in leisure activities outside home | (4) Washing hands | (5) Wearing mask |
|---|---|---|---|---|---|
| Vaccine First | 0.057** | 0.032 | 0.028 | 0.028** | 0.003 |
|  | (0.02) | (0.02) | (0.02) | (0.01) | (0.02) |
| Vaccine second | 0.107*** | 0.090*** | 0.079*** | 0.008 | 0.005 |
|  | (0.02) | (0.03) | (0.02) | (0.01) | (0.02) |
| Adj $R^2$ | 0.52 | 0.66 | 0.37 | 0.62 | 0.49 |
| Obs. | 54,007 | 54,007 | 54,007 | 54,007 | 54,007 |

**Note**: Numbers within parentheses are robust standard errors clustered in the residential prefectures. The set of control variables used in Table 3 is included, although the results are not reported.

***p<0.01

**p<0.05

*p<0.10

## 3.2. Sub-sample estimations (Young vs Old ages groups)

Tables 5 and 6 report the results based on sub-samples below 45 years and sub-samples equal to or over 45 years. Here, we focus on key variables, although the same set of control variables is included.

In contrast to the results in Table 3, Table 5 indicates the negative sign of vaccination dummies for staying indoors and not going out for work. In particular, all dummies for the second shot are statistically significant for the estimations of not going out for work. Furthermore, the absolute values of the coefficients for Vaccine second_1, Vaccine second_2, Vaccine second_3, and Vaccine second_4 are 0.249, 0.392, 0.347, and 0.615, respectively, which suggests that the vaccinated people are more likely to go to work or school than the non-vaccinated ones as time has passed. Moreover, these values were remarkably larger than those of staying indoors. Meanwhile, as for results where not participating in leisure activities outside home is a dependent variable, we did not observe statistical significance in vaccination dummies. Considering the results jointly, vaccinated people in the young group have a stronger motivation to go to work or school than the non-vaccinated ones, whereas they do not have a stronger motivation to go to leisure. We interpreted that they received the shot in the workplace or school and are encouraged or required to go for working or learning.

**Table 5. FE model: Dependent variables are preventive behaviors (ages<=40 years).**

|  | (1) Staying indoors | (2) Not going out for work | (3) Not participating in leisure activities outside home | (4) Washing hands | (5) Wearing mask |
|---|---|---|---|---|---|
| Vaccine First | −0.095 (0.06) | −0.037 (0.06) | −0.025 (0.06) | −0.031 (0.05) | 0.017 (0.03) |
| Vaccine second_1 | −0.106 (0.08) | −0.249*** (0.07) | 0.049 (0.06) | −0.060 (0.05) | 0.021 (0.04) |
| Vaccine second_2 | −0.283** (0.13) | −0.392** (0.10) | 0.029 (0.11) | 0.0002 (0.06) | 0.053 (0.06) |
| Vaccine second_3 | −0.288* (0.15) | −0.347* (0.18) | −0.079 (0.12) | −0.053 (0.12) | 0.142 (0.13) |
| Vaccine second_4 | −0.467 (0.31) | −0.615*** (0.13) | −0.140 (0.29) | 0.004 (0.10) | 0.064 (0.08) |
| Adj $R^2$ | 0.49 | 0.56 | 0.37 | 0.52 | 0.58 |
| Obs. | 15,407 | 15,407 | 15,407 | 15,407 | 15,407 |

Note: Numbers within parentheses are robust standard errors clustered in the residential prefectures. The set of control variables used in Table 3 is included, although the results are not reported.

***$p<0.01$

**$p<0.05$

*$p<0.10$

Switching attention to Table 6, the results of the old generation show a similar r of dummies for vaccination are very similar to those in Table 3. The incentive that young people have is not common for aged groups because observations of the old group consisted largely of retired-aged people. The sample size and observations of vaccinated people for the old group were far larger than those for the younger young group. Hence, the influence of vaccination dummies for the old group outweighs that of the young group, which is reflected in the results of the whole sample in Table 3.

**Table 6. FE model: Dependent variables are preventive behaviors (ages>40 years).**

| | (1) Staying indoors | (2) Not going out for work | (3) Not participating in leisure activities outside home | (4) Washing hands | (5) Wearing mask |
|---|---|---|---|---|---|
| Vaccine First | 0.036 (0.03) | 0.026 (0.03) | 0.013 (0.02) | −0.012 (0.02) | 0.02* (0.01) |
| Vaccine second_1 | 0.089*** (0.03) | 0.105** (0.03) | 0.051* (0.03) | −0.016 (0.02) | −0.001 (0.02) |
| Vaccine second_2 | 0.120*** (0.03) | 0.175*** (0.04) | 0.084*** (0.03) | −0.021 (0.03) | −0.024 (0.02) |
| Vaccine second_3 | 0.118* (0.06) | 0.170*** (0.05) | 0.005 (0.06) | −0.043 (0.03) | 0.012 (0.03) |
| Vaccine second_4 | 0.131 (0.17) | 0.167* (0.09) | −0.123 (0.12) | −0.077 (0.08) | −0.048 (0.08) |
| Adj $R^2$ | 0.52 | 0.68 | 0.35 | 0.53 | 0.63 |
| Obs. | 38,600 | 38,600 | 38,600 | 38,600 | 38,600 |

**Note**: Numbers within parentheses are robust standard errors clustered in the residential prefectures. The set of control variables used in Table 3 is included, although the results are not reported.

***p<0.01

**p<0.05

*p<0.10

# 4. Discussion

In a previous related study conducted in the U.K., individuals' COVID-19 preventive behaviors did not decrease even after vaccination[22,23]. This indicates that preventive behaviors against COVID-19 are motivated by self-interest and other related concerns. This is in line with the expectation that the COVID-19 vaccine protects both the vaccinated individuals and society by reducing the spread of the disease. Moreover, the vaccinated individuals exhibited less generosity toward the non-vaccinated individuals[43].

Using independently collected panel data, we found that the vaccinated individuals are more likely to stay at home, frequently wash their hands, and wear masks than the non-vaccinated ones, consistently from the early stage of COVID-19 and after the distribution of vaccines. The results obtained by analyzing the FE model indicate that the gap between the vaccinated and non-vaccinated individuals in terms of "staying at home" increased. In the context of "washing hands" or "wearing a mask," the gap did not reduce.

Information was diffused through various media to call for a cautious attitude, which possibly formed social norms to engage in preventive behaviors. People displayed preventive behaviors, which depended on caring and fairness concerns because of this norm44. Vaccinated people will be criticized by members of society if they do not take preventive behaviors. This increased the incentive of the vaccinated individuals to take preventive behaviors.

A closer examination finds that young individuals aged equal to or below 40, tend to go out to work post vaccination. They are likely to be vaccinated at their workplace, such that they can work safely. They need to go out to work because working from home is yet to take a firm hold in Japan, and they are less likely to get a post in management that makes working from home possible. Inevitably, the vaccinated young workers seem unlikely to obey the norms. However, apart from going out to work, they continued to display other preventive behaviors, such as refraining from participating in leisure activities outside home, frequently washing hands, and wearing a mask.

Overall, the key findings are consistent with the argument that "individuals act upon the social contract; the stronger they perceive it as a moral obligation, the more they act upon it. Emphasizing the social contract could be a promising intervention to increase vaccine uptake, prevent free riding, and eventually support the elimination of infectious diseases."[43]

## Acknowledgments

We would like to thank Editage (http://www.editage.com) for editing and reviewing this manuscript for the English language.